\documentclass[reprint,showpacs,preprintnumbers,amsmath,amssymb,prc,floatfix]{revtex4-1}
\usepackage{float}
\usepackage{color}
\usepackage{graphicx}
\usepackage{dcolumn}
\usepackage{bm}
\usepackage{xcolor}
\usepackage{multirow}
\usepackage{threeparttable}
\usepackage{adjustbox}
\usepackage{longtable}
\usepackage{comment}
\usepackage{natbib}
\usepackage{array}
\usepackage[utf8x]{inputenc}
\usepackage[colorlinks,citecolor=blue,linkcolor=red,anchorcolor=blue,filecolor=blue,urlcolor=blue]{hyperref}

\begin{document}
\title{Appearance of peak in symmetry energy at N = 126 for Pb isotopic chain within relativistic energy density functional}

\author{Jeet Amrit Pattnaik$^1$}
\email{jeetamritboudh@gmail.com}
\author{T. M. Joshua$^2$}
\email{majekjoe1@gmail.com}
\author{Ankit Kumar$^{3,4}$}
\email{ankit.k@iopb.res.in}
\author{M. Bhuyan$^{5,6}$}
\email{bunuphy@um.edu.my}
\author{S. K. Patra $^{3,4}$}
\email{patra@iopb.res.in}
\affiliation{$^1$Department of Physics, Siksha 'O' Anusandhan, Deemed to be University, Bhubaneswar-751030, India}
\affiliation{$^2$Institute of Engineering Mathematics, Faculty of Applied and Human Sciences, Universiti Malaysia Perlis, Arau, 02600, Perlis, Malaysia}
\affiliation{$^3$Institute of Physics, Sachivalya Marg, Bhubaneswar-751005, India}
\affiliation{$^4$Homi Bhabha National Institute, Training School Complex, Anushakti Nagar, Mumbai 400094, India}
\affiliation{$^5$Center for Theoretical and Computational Physics, Department of Physics, Faculty of Science, University of Malaya, Kuala Lumpur 50603, Malaysia}
\affiliation{$^6$Institute of Research and Development, Duy Tan University, Da Nang 550000, Vietnam}

\date{\today}

\begin{abstract}
\noindent
The newly derived relativistic energy density functional [\textcolor{blue}{ Phys.  Rev. C \textbf{103}, 024305 (2021)}], which stems from the effective field theory motivated relativistic mean-field (E-RMF) is employed to establish the appearance of peak/kink in the symmetry energy over the isotopic chain of Pb- nuclei. The coherent density fluctuation model parametrization procedure for finite nuclei is adopted here to obtain the relativistic energy density functional at local density. The relativistic energy density functional from E-RMFT takes precedence over the Br\"uckner energy density functional as it accurately predicts the empirical saturation density and binding energy per nucleon $E/A$, so-called 'Coester Band Problem'. Interestingly, using the relativistic energy density functional, it is possible to predict the peak at $N=126$ for recently developed G3 and widely used NL3 parameter sets, which is not observed for Br\"uckener's functional in-spite of using the E-RMF density. From the present analysis, the newly fitted energy density functional is found to be minutely sensitive to the choice of the parameter sets employed.  \\
\end{abstract}
\maketitle

\label{intro} 
\noindent
The study of nuclear symmetry energy and its isospin dependence is of central interest in various areas of nuclear physics \cite{niks08,sing14,dale10,chen08} and astrophysical systems including neutron stars \cite{li19,zhan19}. This is one of the most essential characteristics of nuclei that lies far from the stability line. Several recent studies have unequivocally demonstrated that the presence of kinks in symmetry energy of finite nuclei over the isotopic chain points to the existence of magic numbers \cite{anto16,bhuy18,kaur20a,kaur20b,qudd20,patt21}. The  advancement in the sophisticated experimental radioactive ion beam facilities 
using the projectile-fragmentation reaction method with fast ion-beams
at FRIB (USA) \cite{thoe10}, RIKEN (Japan) \cite{saku08}, GSI (Germany) \cite{geis92}, FLNR (Russia) \cite{rodi03}, CSR (China) \cite{sun03} and SPIRAL2/GANIL (France) \cite{muel91} have led to the renaissance of investigating the behaviour of symmetry energy at and above the saturation density \cite{li08}. An alternative way to produce the radioactive ion beam is the isotope separator on-line (ISOL) technique has recently gained momentum for high asymmetric iso-spin nuclei \cite{tohr14}. The notable examples among them are at Lou-vain la Neuve (Belgium)\cite{dec91}, Spiral(France)\cite{bou14}, Alto (France)\cite{fran15,essa13}, ISAC (Canada)\cite{reit20} and REX ISOLDE (Switzerland/France) \cite{fras17}. In this context, the symmetry energy is closely associated with the isospin asymmetry of both finite and infinite nuclear matter, and hence, serves as a bridge between them \cite{van10}. To predict this isospin and density-dependent quantity, so-called symmetry energy for both infinite nuclear matter and finite nuclei, one has to use a reliable and convenient approach that works a wide range of density, i.e., from sub- to supra-saturation density.   

Conventionally, various approaches such as the liquid drop model \cite{myer66,pomo03}, Skyrme energy density functional (SEDF) \cite{chen10,dutr12}, Hartree-Fock with random phase approximation (RPA) \cite{carb10},  relativistic mean-field with RPA predicated on effective Lagrangians with density-dependent meson-nucleon vertex functions \cite{vret03} among other many-body approaches, have been employed to study the symmetry energy of finite nuclei at a local density approximation. In parallel to these the well known Br\"uckner energy density functional (EDF) \cite{brue68,brue69} within the coherent density fluctuation model (CDFM) as those in Refs. \cite{bhuy18,qudd20,gaid11,gaid12} is successfully applied for the study of surface properties of nuclei. It is well-known that the classical Br\"uckner energy density functional fails to satisfy the Coester-Band problem \cite{coes70,brock90} i.e. inability to accurately reproduce the empirical saturation density $\rho\approx0.15$ fm$^{-3}$ and binding energy per nucleon in the limit $E/A\approx-16$ MeV \cite{myer96}. Later on, a few alternative attempts which considered the incorporation of various realistic nucleon-nucleon potentials \cite{coes70,day81} into the Br\"uckner energy density functional are found either to overestimate the nuclear matter density or underestimate the binding energy. In other words, the saturation properties must satisfy the nuclear equation of state (EoS) while extrapolating to higher density and isospin asymmetry.

Recently, the surface properties of nuclei can be estimated by using the non-relativistic and relativistic inputs within the coherent density fluctuation model (CDFM) \cite{bhuy18,qudd20,patt21,gaid11,gaid12}. From these analysis, one can find the notable peaks/kinks that have been observed for traditional magic neutron and/or proton. For example, $N = 20, 28$ for Ca and $N=82$ for Sn  along with a few predictions for drip-line and superheavy island  \cite{kaur20b,bhuy18,qudd20,patt21,gaid11,gaid12}. On the other hand, these studies are unable to reproduce peak/kinks for Pb $(N=126)$, which may be associated with the saturation properties. In this context, it necessitates a well-grounded approach that can both tackle the Coester-Band problem \cite{coes70,brock90} as well as the supra-saturation object including neutron star. In our previous work, a newly fitted relativistic density functional has been introduced \cite{kuma21} and tested for a few double magic nuclei. Hence, it is our main objective to test the newly relativistic energy density functional for the isotopic chain of Pb nuclei. In the present analysis, the symmetry energy and {its volume and surface components} are calculated by using the relativistic energy density functional (EDF), which stems from the effective-field theory motivated relativistic mean-field (E-RMF) model  \cite{kumar17,kumar18} for the well-known NL3 and recently developed G3 parameter sets. 
\\
\\
The zeroth component of the energy-momentum tensor $T_{00}$ gives the energy density of the system as a function of scalar and vector density $\rho_s$ and $\rho_v$ respectively\cite{kuma21}:
\begin{eqnarray}
{\cal{E}}(k)_{nucl.}&=&\frac{2}{(2\pi)^{3}}\int d^{3}k E_{i}^\ast (k)+
\frac{ m_{s}^2\Phi^{2}}{g_{s}^2}\Bigg(\frac{1}{2}+\frac{\kappa_{3}}{3!}
\frac{\Phi }{M} 
\nonumber\\
&&
+ \frac{\kappa_4}{4!}\frac{\Phi^2}{M^2}\Bigg)
+\rho_b W-\frac{1}{4!}\frac{\zeta_{0}W^{4}}
{g_{\omega}^2}
-\frac{1}{2}m_{\omega}^2\frac{W^{2}}{g_{\omega}^2}
\nonumber\\
&&
\Bigg(1+\eta_{1}\frac{\Phi}{M}+\frac{\eta_{2}}{2}\frac{\Phi ^2}{M^2}\Bigg)
+\frac{1}{2}\rho_{3}R
-\frac{1}{2}\Bigg(1+\frac{\eta_{\rho}\Phi}{M}\Bigg)
\nonumber\\
&&
\frac{m_{\rho}^2}{g_{\rho}^2}R^{2}
-\Lambda_{\omega}  (R^{2}\times W^{2})
+\frac{1}{2}\frac{m_{\delta}^2}{g_{\delta}^{2}}D^{2}.
\label{enm}
\end{eqnarray}

The conversion of the nuclear matter quantities [Eq. (\ref{enm})] from momentum space to the coordinate space is the major distinctive of the calculation. In other words, the nuclear matter (NM) quantities are reconstructed at local density. It is assumed that the NM is made up of tiny spherical pieces described by a local density function termed as {\it Flucton} defined as $\rho_0(x) = 3 A/4\pi x^3$. As such, the fitted binding energy function of E-RMF is expressed as \cite{kuma21}:
\begin{eqnarray}
{\cal E}(x) & = & C_k \rho_0^{2/3}(x) + \sum_{i=3}^{14} (b_i + a_i \alpha^2) \rho_0^{i/3}(x).
\label{efitting}
\end{eqnarray}
 The first term represents the kinetic energy, whose coefficient $C_k$ is given as $C_k = 37.53 [(1+\alpha)^{5/3} + (1-\alpha)^{5/3}]$ following the Thomas-Fermi approach. The asymmetry parameter $\alpha$ is defined as $\alpha = \frac{\rho_n-\rho_p}{\rho_n+\rho_p}$, with  $\rho_n$ and $\rho_p$ are the neutron and proton densities distribution, respectively. A number of terms are involved in the polynomial fitting  (Eq. (\ref{efitting})) which is used to obtain the exact nature of the binding per particle $E/A$ in position space. The mean deviation is calculated using the formula $\delta=\Big[\sum_{j=1}^{n} (E/A)_{j,Fitted} - (E/A)_{j,RMF}\Big]/n $. The term $(E/A)_{j, Fitted}$ represents the binding energy deduced from Eq. (\ref{efitting}),  $(E/A)_{j, RMF}$ is the binding energy per nucleon from the RMF functional and $n$ is the number of data points. From the analysis \cite{kuma21}, it is found that 12 terms are considered to obtain the best fit  and the coefficients extrapolated from the polynomial fitting are presented in Table -I of the Ref. \cite{kuma21}. \\
The nuclear matter symmetry energy $S^{NM}$ is obtained from the following standard relations \cite{kumar18,Chen14}

\begin{eqnarray}
S^{NM}&=&\frac{1}{2}\frac{\partial^2 ({\cal E}/\rho)}{\partial\alpha^2}\Big|_{\alpha=0}.\label{snm}
\end{eqnarray}
which is given as follow using Eqs. (2-3).
\begin{eqnarray}
S^{NM} &=& 41.7\,\rho_0^{2/3}(x) + \sum_{i=3}^{14} a_i\, \rho_0^{i/3}(x), \label{eqB}
\end{eqnarray}

\begin{figure}[H]
\centerline{\includegraphics[width=1.05 \columnwidth]{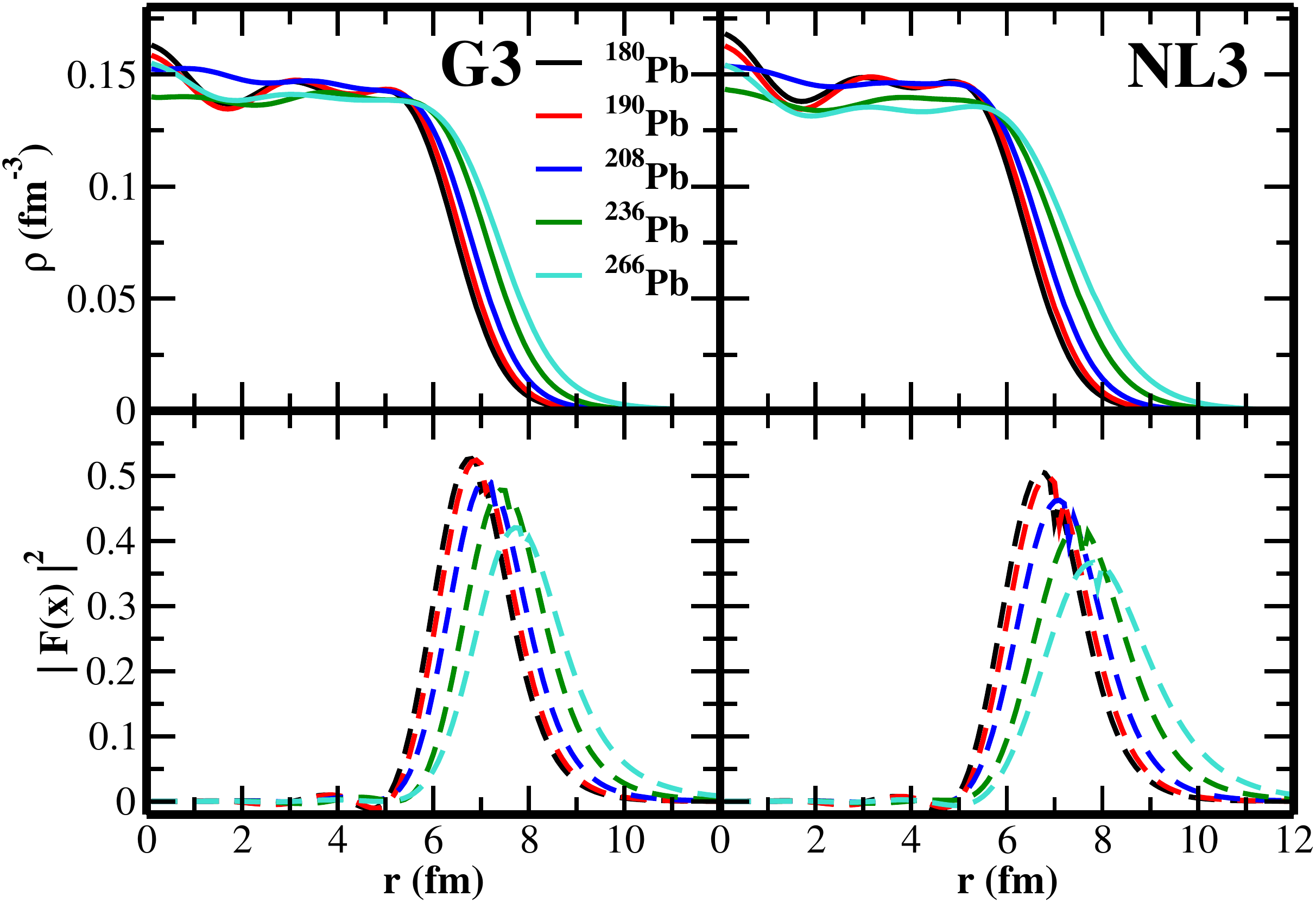}}
\vspace*{8pt}
\caption{(Color online) The E-RMF densities and weight functions for $^{180,190,208,236,266}$Pb with G3 and NL3 parameter sets. The solid line (upper panel) is the density and the dashed line (lower panel) is the weight function $\vert F(x) \vert^2 $ for the considered nuclei.} \protect\label{fig3}
\end{figure}

The density of closed/semi-closed-shell spherical Pb nucleus is calculated using E-RMF formalism and used as input in the CDFM to calculate the weight function, which is the major quantity that bridges nuclear matter (NM) parameters in $x-$space and finite nuclei in $r-$space using local density approximation (LDA). The $r-$ and $x-$space are matched together,  by the superposition of the total density of the nucleus  and an infinite number of $Fluctons$, using the coherent density fluctuation model approach. 
\\
\\
The expression for the energy density of infinite and isotropic NM are obtained from the Br\"{u}ckner functional defined as \cite{brue68,brue69}:
\begin{eqnarray}
{\cal{E}}(\rho)_{nucl.}&=&AV_0(x)+V_C-V_{Cx},
\label{brue0}
\end{eqnarray}
where
\begin{eqnarray}
V_0(x)&=& 37.53\Big[(1+\alpha)^{5/3} + (1-\alpha)^{5/3}]\rho_0(x)^{2/3} \nonumber \\ 
&& + b_1\rho_0(x) +b_2\rho_0(x)^{4/3} + b_3\rho_0(x)^{5/3} \nonumber \\
&& +\alpha^2[b_4\rho_0(x) + b_5\rho_0(x)^{4/3} + b_6 \rho_0(x)^{5/3} \Big].
\end{eqnarray}
Here, $b_1= -741.28$, $b_2=1179.89$, $b_3=-467.54$, $b_4=148.26$,
$b_5=372.84$, $b_6=-769.57$ and the total density $\rho = \rho_n+\rho_p$ is the sum of the neutrons and protons density distributions\cite{kaur20a}. In each {\it  Flucton} there are protons having Coulomb energy $V_C=\frac{3}{5}\frac{Z^2e^2}{x}$
and Coulomb exchange energy $V_{Cx}=0.7386Ze^2(3Z/4\pi x^3)^{1/3}$.
The important part of the present calculation is to convert the nuclear matter quantities Eq. (\ref{brue0}) from momentum ($\rho-$) to coordinate ($r-$) space in local density approximation (LDA). The nuclear matter symmetry energy parameter $S^{NM}$ is obtained from the well defined relation \cite{kumar18,Chen14,gaid11}:
\begin{eqnarray}
S^{NM}&=&\frac{1}{2}\frac{\partial^2 ({\cal E}/\rho)}{\partial\alpha^2}\Big|_{\alpha=0}
\nonumber
\\
&=& 41.7\rho_0(x)^{2/3}+b_4\rho_0(x)+b_5\rho_0(x)^{4/3}
+ b_6\rho_0(x)^{5/3}
\nonumber
\label{snm} \\ 
\end{eqnarray}
 at local density.
 The weight function $|F(x)|^2$ for a given density $\rho$ (r) is defined as
\begin{equation}
|F(x)|^2 = - \left (\frac{1}{\rho_0 (x)} \frac{d\rho (r)}{dr}\right)_{r=x},
\label{wfn}
\end{equation}
with $\int_0^{\infty} dx \vert F(x) \vert^2 =1$. More comprehensive analytical derivation are found in Refs. \cite{bhuy18,gaid11,gaid12,anto80,anto82}. Here the E-RMF densities obtained from G3 and NL3 parameter sets are used as local density to calculate the weight function as defined in 
Eq. (\ref{wfn}). The densities near the surface region monotonically decreases for finite nuclei, which produces a peak in the weight function for this region. It is to be noted that the peak of the weight function appears in the tail part of the density distribution. Further discussion is given below in the result section with Fig. \ref{fig3}. The coherent density fluctuation model CDFM provides an easy transition from the properties of nuclear matter to those of finite nuclei. The finite nucleus symmetry energy $S^{A}$ with mass number A is calculated by weighting the corresponding quantity for infinite nuclear matter within the CDFM, as given below \cite{gaid11,gaid12,anto17}
\begin{eqnarray}
S^{A}= \int_0^{\infty} dx\, \vert F(x) \vert^2\, S^{NM} (\rho (x)) ,
\label{s0}
\end{eqnarray}
The symmetry energy $S^{A}$, in  Eq. ($\ref{s0}$) is the surface weighted average of the corresponding nuclear matter quantity in the local density approximation LDA limit for finite nuclei. To estimate the symmetry energy, the densities are obtained self-consistently from E-RMF and folded with the nuclear matter parameters using the CDFM.
\\
\\
The surface $S^A_S$ and volume $S^A_V$ components of symmetry energy are analysed separately in the frame-work of Danielewicz's Liquid Drop model \cite{dani03,dani04,dani06}. The symmetry energy $S^A$ is connected with the surface and volume components as \cite{steiner2005}:
\begin{eqnarray}
S^A= \frac{S^A_{V}}{1+ \frac {S^A_{S}} {S^A_{V}} A^{-1/3}}= \frac{S^A_{V}}{1+A^{-1/3}/\kappa}.
\label{eq4}
\end{eqnarray}
From  Eq. (\ref{eq4}), the individual components of $S^A_V$ and $S^A_S$ can be written as:
\begin{eqnarray}
S^A_{V}=  S^A \left (1+\frac{1}{\kappa A^{1/3}} \right)
\end{eqnarray}
and
\begin{eqnarray}
S^A_{S}= \frac {S^A}{\kappa} \left (1+\frac{1}{\kappa A^{1/3}} \right).
\end{eqnarray}
Here, $\kappa \equiv \frac{S^A_{V}}{S^A_{S}}$ is the ratio of the volume and surface symmetry energy. The symmetry energy and its volume and surface components are calculated within the CDFM formalism \cite{bhuy18,dani03,dani04,dani06,kaur20a,steiner2005,dan20,gaid21}. Recently an alternative approach has been introduced by Gaidarov \textit{et al.} to obtain the volume and surface symmetry energy components by taking the non-relativistic densities in the weight function \cite{gaid21}. 
\\
\\
Fig. \ref{fig3} represents the densities of 3 different regions of Pb- isotopic series. We have considered the densities of $^{180}$Pb, $^{190}$Pb, $^{208}$Pb, $^{236}$Pb and $^{266}$Pb for the $\beta$−stable and drip-line regions as representative cases ranging from the neutron deficient, $\beta-$stable and neutron-rich regions, which analyses the relative changes of the density with respect to neutron-proton isospin asymmetry. From the E-RMF density, we acquire the weight function $\vert F(x) \vert^2 $ of the nucleus. Here, $\vert F(x) \vert^2 $ is one of the crucial factors to determine the surface properties of the nucleus. This value is significant with a radial distribution of $\sim 5-10$ fm for $^{180,190,208}$Pb and for $^{236,266}$Pb the crucial range is $\sim 5-12$ fm showing the surface properties of the nuclei (see lower panel of Fig. \ref{fig3}). Because, this region of the nucleus lies in the tail part of the density distribution of nucleons as shown in the upper panel of Fig. \ref{fig3}. 
\begin{figure}
\centerline{\includegraphics[width=1.05 \columnwidth]{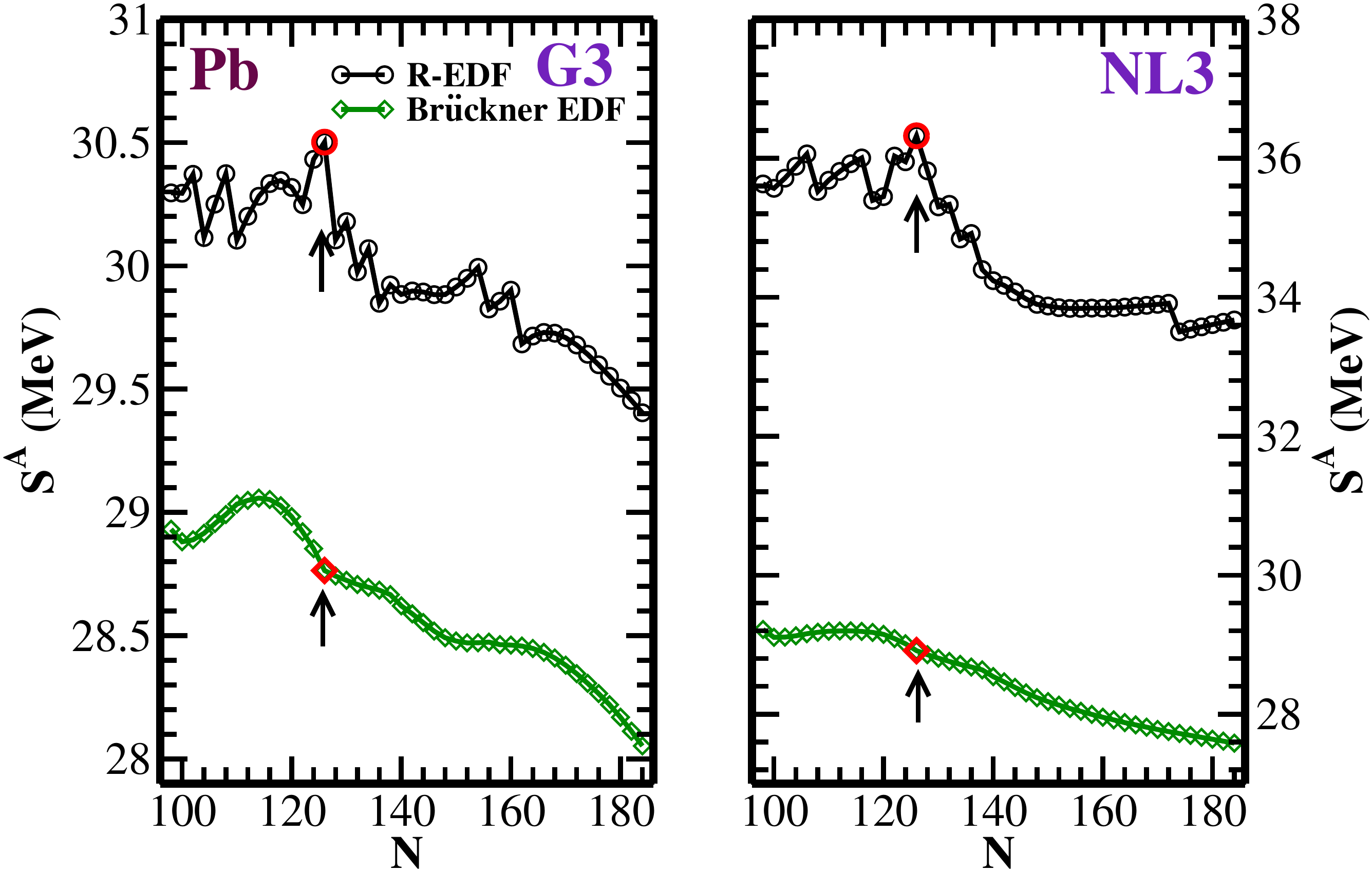}}
\vspace*{8pt}
\caption{(Color online) The nuclear symmetry energies are shown for the relativistic energy density and Br\"uckner functionals for the Pb isotopes with NL3 and G3 parameter sets. The arrow represents the value of $S^A$ at $N=126$. \protect\label{fig1}}
\end{figure}

The peak in the surface properties such as nuclear symmetry energy and its component are used as indicators of shell/sub-shell closure \cite{bhuy18,qudd20,patt21,gaid11,gaid12}. The possible reasons for the disappearance of the peak at neutron number N=126 for Pb nuclei have been discussed. It is reported that the peak shifted a few units (N=120) than N=126 when the Br\"uckner energy density functional is used in the evaluation of symmetry energy within the coherent density fluctuation model formalism. The use of the self-consistent E-RMF density does not improve the situation much. To see the effect on the functional chosen, here, the recently developed relativistic energy density functional (EDF) \cite{kuma21}, which systematically incorporates the Coester-Band problem \cite{coes70,brock90} via the RMF Lagrangian density is compared to the conventional Br\"uckner functional.
\\
\\
Figure \ref{fig1} displays the profile of the symmetry energy $S^A$ as a function of  neutron numbers ranging from N=98 to N=184 for Pb-isotopes. Here, only the even-even nuclei are considered to preserve the time-reversal symmetry and also avoid the odd-even staggering. A comparison is made between the symmetry energy obtained from Br\"uckner (green diamonds) and relativistic EDFs (black open-circle)  to assert their respective suitability in the Pb isotopic chain. Although the trends seem to be somewhat similar, both predictions fall between  different ranges of values. In other words, Br\"uckener's predicted values are bounded between 28.0 MeV and 29.2 MeV while those from RMF with G3 force parameter are found between 29.4 MeV and 30.6 MeV over the isotopic chain.
Similarly, for NL3 set, the Br\"uckner's prediction is in the range 27.5-29.2 MeV and the R-EDF result is in between 33.6-35.7 MeV for the considered Pb chain. The symmetry energy at saturation for nuclear matter with G3 and NL3 forces are 31.8 and 37.4 MeV, respectively \cite{kumar17,kumar18}.
At N=126, the symmetry energy of the relativistic energy density functional differs by 1.73 MeV for G3  from it's Br\"uckner prediction, whereas this difference for NL3 is 7.40 MeV. This implies the E-RMF approach gives larger symmetry energy than the Br\"uckner approach. Further, we noticed few maxima and minima in the E-RMF prediction of symmetry energy both in G3 and NL3 models indicating the structure effect of the isotopic chain. However, it is to be noted that these variations in $S^A$ are very small and can not be considered as the shell/sub-shell closure. In the case of NL3, few other maxima appears in the symmetry energy curve. Although these peaks almost comparable with N=126, we do not consider those as shell/sub-shell closures because of the small fluctuation in the values of $S^A$. Mostly we respect the trend of the curve while locating the shell/sub-shell closure as shown in the Fig.\ref{fig1}. Furthermore, from Br\"uckner's prescription, no significant peak is found at $N=126$ shell closure, corresponding to $^{208}$Pb as also reported in earlier studies \cite{bhuy18,qudd20,patt21,gaid11,gaid12} and the references therein. This peak shifted to N=120 which is not a close shell for Pb nucleus. In contrast, interestingly, the relativistic EDF shows a notable peak at the neutron shell closure $N=126$ in G3 and NL3 parameter sets. 
More careful inspection shows that the trajectories of the symmetry energy exhibit an anomalous trend in both cases, which is very common in the mean-field calculation due to the structure effect of the nucleus. Although the predictions for both the parameter sets have different magnitude, it is obvious that they are characterised with the same trends for the shell/sub-shell closure. This indicates that the relativistic energy density functional predictions for shell/sub-shell closure within surface properties such as symmetry energy and its components are merely sensitive to the choice parameter set used.\\ 
\\
\begin{figure}[H]
\centerline{\includegraphics[width=1.05 \columnwidth]{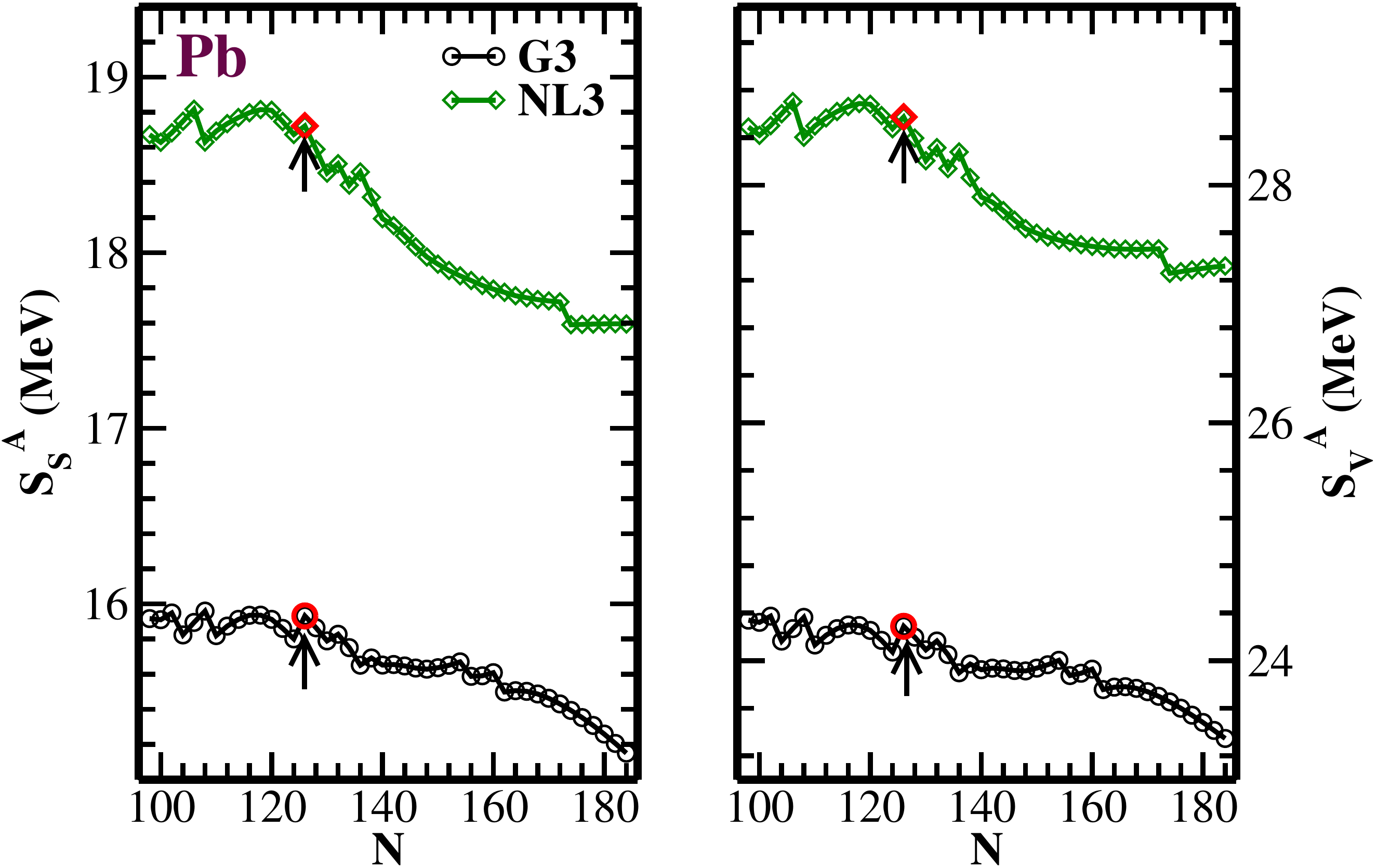}}
\vspace*{8pt}
\caption{(Color online) The surface ($S_S^A$) and volume ($S_V^A$) symmetry energies are estimated using Danielewicz's liquid drop prescription for the Pb isotopes with NL3 and G3 parameter sets. The arrow represents the symmetry energy at neutron number $N=126$.\protect\label{fig2}}
\end{figure}
Fig. \ref{fig2} displays the variation of  $S^A_S$ and $S^A_V$ over the Pb-isotopic chain as a function of the neutron number N. There exist a strong correlation between the parameters $S^A_S$ and $S^A_V$. Although a similar trend is observed in their profiles, it is clear that their respective magnitudes vary with their neutron numbers and they are bound between different energy ranges. Particularly, from their respective values, the volume component contributes more to the symmetry energy. In both cases, a notable peak/kink is found at the neutron shell closure N=126. Again, the NL3 predictions seem to be characterized with lesser undulations/fluctuations as compared to the G3 parameter set. One of the striking observations in the relativistic energy density functional framework is that, parametrization plays a key role in its calculations. More elaborate details on the surface and volume components of the symmetry energy as well as their parametrizations and respective contributions can be found in Refs. \cite{agra12,satu06}.
\begin{figure}[H]
\centering
\centerline{\includegraphics[width=1.02 \columnwidth]{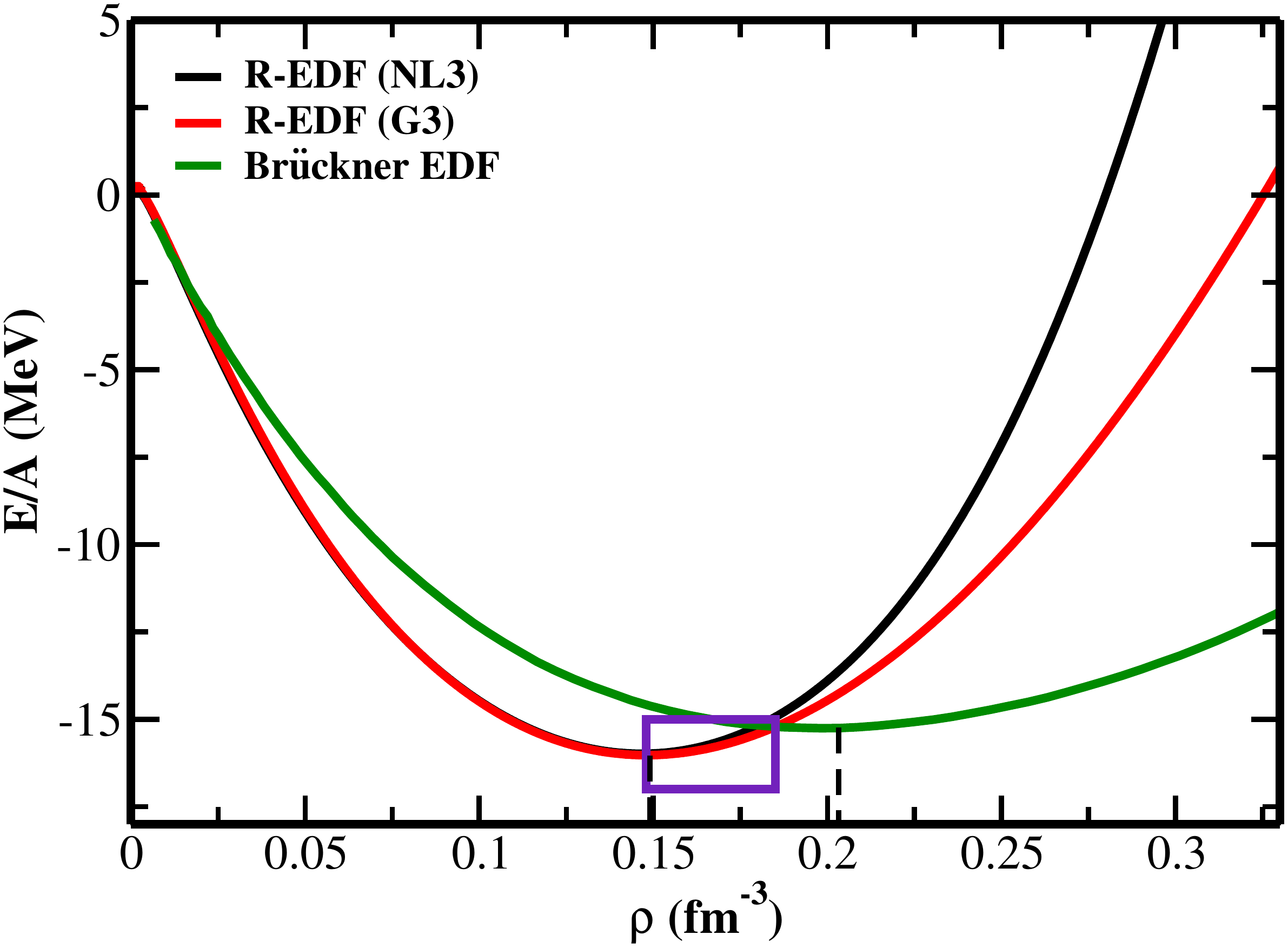}}
\caption{(Color online) Binding energy per particle ($E/A$) as a function of nuclear matter density ($\rho$) for NL3, G3 and Br\"ueckner energy density functionals. The rectangular box indicates the Coester band with empirical density and binding energy per nucleon \cite{coes70,brock90}.  \protect\label{fig4} }
\end{figure}


Having obtained a peak at $N=126$ using the relativistic energy density functional  with the G3 and NL3 parameter sets, it is yet necessary to validate the appropriateness of this functional in the wake of the Coester-Band problem. Figure \ref{fig4} shows the variation of the nuclear matter binding energy per nucleon $E/A$  as function of the baryonic density for symmetric nuclear matter with the Br\"uckner (green solid line) and RMF (G3  and NL3 with red and black solid lines respectively) functionals. Relatively, one can vividly notice that while $E/A$ shifts towards the lower density region, whereas the deepest minimum of in the Br\"uckner's prescription turns out to underestimate $E/A$  as -14.9 MeV and overestimate the saturation density $\rho$ to be $\sim$ 0.2 fm$^{-3}$. In case of relativistic energy density functional, the deepest minimum passes through the band at $E/A=-16$ MeV and $\rho=0.15$ fm$^{-3}$  than the empirical one as shown in the rectangular box \cite{coes70,brock90}.  This resolves the issue of the Coester band problem \cite{coes70,brock90}, which is directly connected with the energy density, and the isospin dependant quantities such as symmetry energy and its component and/or co-efficient. Hence, the dilute picture of peak at the magic number in the case of Br\"uckener's functional can be correlated with the Coester band problem. In other words, the successful achievement of the Coester band problem within relativistic energy density functional is reflected in the calculation of symmetry energy and its component and reproduce the peak of Pb isotopes at N = 126 contrary to Br\"uckner's functional at N=120. This shifting of the peak from N=126 to N=120 may be correlated with the shifting of the energy minimum to the Fermi momentum value $k_f\sim 2.0 fm^{-3}$.
\\

The surface properties of Pb isotopes are evaluated using the newly fitted expression of Ankit \textit{et al.} derived from E-RMF formalism to resolve the Coester-Band problem encountered in the conventional Br\"uckner energy density functional. The E-RMF densities with the NL3 and G3 parameter sets are used as inputs to  obtain the weight function within the coherent density fluctuation model. The relativistic energy density  functional manifests a remarkable success in accurately reproducing the empirical saturation density $\rho$ as well as the binding energy per nucleon $E/A$. Hence, it is established that this approach has an edge over the Br\"uckner's prescription.

Besides, the  relativistic energy density functional (R-EDF) is used to establish the existence of peak in the symmetry energy of finite nuclei over the Pb isotopic chain at $N = 126$ corresponding to the double magic $^{208}$Pb nucleus which has been hitherto elusive within the conventional Br\"uckner's prescription. Although, the trend in the Pb-isotopic chain is not smooth enough (especially with the G3 parameter set), a careful inspection reveals that the relativistic EDF  is susceptible to the choice of parameter set and may be a manifestation of the structure effects of the nucleus. In other words,  a smoother behaviour is observed using the NL3 parameter set. This infers that the result could be improved by choosing an appropriate parameter set and taking into account the proper structure effects.
\\
\\
\noindent 
{\bf  Acknowledgement:} One of the authors (JAP) is thankful to the Institute of Physics, Bhubaneswar, for providing computer facilities during the work. SERB partly reinforces this work, Department of Science and Technology, Govt. of India, Project No. CRG/2019/002691. MB acknowledges the support from FOSTECT Project No. FOSTECT.2019B.04, FAPESP Project No. 2017/05660-0, and the CNPq - Brasil. TMJ acknowledges the support from the Fundamental Research Grant Scheme (FRGS) under the grant number FRGS/1/2019/STG02/UNIMAP/02/2 from the Ministry of Education Malaysia stipulated with the Institute of Engineering Mathematics (IMK) of the Faculty of Applied and Human Sciences, UniMAP as the beholder.\\

\end{document}